# Enhancing science instruction in the elementary schools

Walk into an elementary school classroom and watch the excitement of children learning science. Contrast this experience with that of a physics professor standing in front of a sea of first-year students in an introductory college physics course. The first group of students is often enthusiastic about the subject matter and cares about what the teacher is saying. In contrast, the second group consists mostly of students trying to expend the minimum effort needed to achieve the maximum possible grade. The college students are often frustrated by the work required and resentful of teachers who do not recognize that a high grade is their principal goal.

What causes the enthusiasm of the first group to dwindle? Is this change an inevitable consequence of adolescence? Is it a result of bad teaching? Has this loss of interest been present at all times and places, or is it a recent phenomenon that mainly occurs at big state colleges and universities? Is it exacerbated by grade inflation and/or over-emphasis on grades? Is the problem an ironic consequence of society's concern about quality teaching, leading to student evaluations and "performance standards," which have had counterproductive effects (pandering, grade inflation, teaching-to-the-test)? Is it worse for physics, or science in general, than other disciplines? These are difficult questions that lack simple answers.

Problems with student interest and achievement in American science education have been recognized for a long time. One of us (Cole) was in high school when Sputnik was launched, precipitating hand-wringing concern about U.S. science education, which resulted in some changes in the science curriculum and support. Subsequently, these problems have not gone away. For example, recent U.S. student performance in international mathematics and science competitions has been disappointing. The most recent (1999) TIMSS-R comparative study of eighth grade students' understanding of math and science found that the average performance of students in the U.S. is below the average of students in the 23 countries participating in this assessment.[1] The overall performance of U.S. students was below that of students from two similar immigrant countries, Australia and Canada, but about the same as students in Great Britain. Most importantly, our students' average performance on these science and mathematics exams declines (relative to students elsewhere) with grade level. To deal with this problem is a serious concern. A National Academy of Sciences report[2] affirms that "The challenge [to improve science education] extends to everyone within the system. Efforts will be time-consuming, expensive, and sometimes uncomfortable. They will also be exhilarating and deeply rewarding. There is no more important task before us as a nation." In particular, this report encourages "… scientists and engineers to work with school personnel to initiate and sustain the improvement of school science programs."[3]

What can we, as individuals, do? In the following, we discuss some recent collaborative activities undertaken by the two of us, a physicist and a science teacher educator, aimed at enhancing elementary school science teaching and learning. These activities exemplify the kinds of projects that might be useful when synergistic working relationships are established.[4]

Our work is an extension of an ongoing partnership between Penn State University and the State College Area School District.[5] Our collaboration, which we nicknamed the "flight team," consists of two elementary teachers, a curriculum support teacher, a volunteer pilot from US Airways, and the two of us. Our goal is to enhance the Air and Aviation unit, which is part of the 3rd and 4th grade curriculum in our local schools. We met regularly during the spring of 2001 and engaged in extensive discussions, aimed at clarifying concepts involved in flight and crafting opportunities for children to participate in elements of scientific inquiry. The emphasis on meaningful science learning and scientific inquiry is fundamental to contemporary reform efforts in science education.[6]

Our discussions were a learning experience for all of us. The emphasis on flight concepts fit well with our focus on teaching science as inquiry. Given that the physical principles governing our world are often counterintuitive, thinking about them raises interesting, testable questions. To demonstrate the concept of lift and its application to flight, the teachers on our team suggested that a wind tunnel would be a valuable demonstration device. A very talented graduate student in physics at Penn State, John Huckans, succeeded in building a wind tunnel at a modest cost. The students construct airfoils of their own design and test them in the wind tunnel. They can then make changes to their design based on their observations during the initial trials.[7] This project has provoked much discussion among the participants, which now includes two undergraduate students, a physics major and a prospective science teacher, and we have had ongoing debates about the importance of the Bernoulli force in really keeping planes in the air.

Two other projects are underway. One is the production of a video aimed at explaining flight principles and practice to students. Video clips from the longer instructional video will be available on the project web site, <http://www.ed.psu.edu/pds>. The second project is the development of an online tool, Faculty Forum. The Faculty Forum began as a database of unit-related web sites specific to the local elementary school curriculum. Web sites are added to the database by teachers who have reviewed them and/or have used them to enhance the teaching of particular units. Teaching tips and other information are included. Although this resource was originally targeted at science units that lack useful resources, it has been extended to include web sites for other aspects of the elementary curriculum. In addition, two components have been added to the Faculty Forum – a library of Technology Enhanced Lessons and an electronic bulletin board for teachers to discuss issues and share ideas.

Not only have these projects reached local teachers and their children, but they are also influencing the development of the next generation of teachers. The science educator on the flight team, Zembal-Saul, has modified her science methods course for elementary education majors to include model lessons adapted from the enhancements to the Air and Aviation unit. Prospective teachers experience first-hand what it is like to interact with phenomena associated with flight, ask testable questions, collect data, examine their data for patterns, and construct evidence-based explanations. In other words, they experience what it is like to learn science concepts by engaging in scientific inquiry, which is in stark contrast to the ways in which they report having learned science previously. Ultimately, the prospective teachers draw upon what they have learned about supporting students' meaningful science learning and scientific inquiry to design and teach inquiry-based science lessons to children.

From this process, the physicist on the flight team, Cole, has learned much about the physics of flight; more importantly, he has been reminded of the difficulty of learning new concepts and jettisoning old ones. A most rewarding aspect of this activity has been the opportunity to meet and learn from other science educators about their challenges, which are frequently different and more difficult than those encountered in teaching a university physics course.

By our activities we and others have made small steps toward addressing the problems of science education mentioned at the beginning of this editorial. There are many components of this process: making science teachers feel comfortable with and enthusiastic about science is one in which we are particularly involved. No less important is the need to revise college science courses, which is the most direct way that physics faculty can improve the appreciation and understanding of science.[8]

Milton W. Cole (Department of Physics, Eberly College of Science)
Carla Zembal-Saul (Department of Curriculum and Instruction, College of Education)
Penn State University
University Park, PA 16802